\documentclass[a4paper,11pt,english]{article}
\usepackage{geometry}
\usepackage[T1]{fontenc}
\usepackage{babel}
\usepackage{graphicx}
\usepackage{caption,subcaption}
\usepackage{bm}
\usepackage{amsmath,amssymb,amsthm}
\usepackage{url,hyperref}
\usepackage{jabbrv,doi}
\hypersetup{pdfborder = 0 0 0}
\usepackage{authblk}

\newif\ifsubmit
\submittrue

\ifsubmit
\usepackage{tikzexternal}
\else
\usepackage{tikz,tikzscale,pgfplots}
\usetikzlibrary{external}
\fi
\let\oldincludegraphics\includegraphics
\renewcommand{\includegraphics}[2][]{\tikzsetnextfilename{#2}\oldincludegraphics[#1]{#2}}  
\renewcommand{\v}[1]{\bm{#1}}
\renewcommand{\d}{\mathrm{d}}
\DeclareMathOperator{\erfc}{erfc}

\newcommand{\avg}[1]{\left< #1 \right>}

\title{Ewald summation for the rotlet singularity of Stokes flow} 

\author{Ludvig af Klinteberg\thanks{Email address:
    \href{mailto:ludvigak@kth.se}{ludvigak@kth.se}} }

\affil{
  Numerical Analysis, Department of Mathematics,\\
  KTH Royal Institute of Technology, 100 44 Stockholm, Sweden }

\date{}

\begin{document}
\maketitle

\begin{abstract}
  Ewald summation is an efficient method for computing the periodic sums that appear when
  considering the Green's functions of Stokes flow together with periodic boundary
  conditions. We show how Ewald summation, and accompanying truncation error estimates,
  can be easily derived for the rotlet, by considering it as a superposition of
  electrostatic force calculations.
\end{abstract}

\section{Introduction}

The fundamental free-space singularities of Stokes flow are (see
e.g. \cite{Pozrikidis1992}) the stokeslet $S$, the stresslet $T$ and the rotlet
$\Omega$. They are defined (up to a constant) as
\begin{align}
  S_{jl}(\v r) &= \frac{\delta_{jl}}{r} + \frac{r_jr_l}{r^3}, \\
  T_{jlm}(\v r) &= \frac{r_jr_lr_m}{r^5}, \\
  \Omega_{jl}(\v r) &= \epsilon_{jlm}\frac{r_m}{r^3} .
\end{align}
These singularities are central when solving Stokes' equation using boundary integral
methods \cite{Pozrikidis1992}. In the context of flow simulations it is common to use
periodic boundary conditions \cite{AfKlinteberg2014a}, in which case periodic sums of the
above singularities must be considered. Due to the relatively slow decay of the
singularities with respect to distance, some kind of special method is required for
this. A well-established alternative is that of Ewald summation, which has its roots in
electrostatic lattice calculations. It was derived by P.P. Ewald \cite{Ewald1921}, and has
as its central idea to split the kernel of the summation into one short-range component
and one long-range component (for an introduction see e.g. \cite{Deserno1998}). To use
Ewald summation for a given kernel function, one must first derive an Ewald decomposition of
it. Such decompositions are available in the literature for the stokeslet
\cite{Hasimoto1959,Pozrikidis1996} and the stresslet \cite{Fan1998}. For the rotlet, a
decomposition can be found in \cite{Maboudi2014}.

We will here show how a decomposition for the rotlet, which in the end is identical to
that in \cite{Maboudi2014}, can be derived by drawing a parallel to Ewald summation for
the electrostatic force potential. Not only does this parallel give us a shortcut for
deriving the decomposition, it also allows us to derive truncation error estimates by
using results which are well-known in the context of electrostatics.

\section{Rotlet sum in free space}
We consider the rotlet defined as
\begin{align}
  \Omega_{jl}(\v r) = \epsilon_{jlm}\frac{r_m}{r^3} .
\end{align}
For a set of $N$ point sources $\v f^n$ at locations $\v x^n \in \mathbb R^3$, the
corresponding velocity field (which we will also refer to as the rotlet potential) at a
target point $\v x$ is
\begin{align}
  u_j(\v x) &= \sum_{n=1}^N \Omega_{jl}(\v x-\v x^n) f^n_l\\
  &= \sum_{n=1}^N\epsilon_{jlm}\frac{x_m- x_m^n}{|\v x- \v x^n|^3} f_l^n .
\end{align}
Recognizing that the kernel $\v r/r^3$ is also used for electrostatic force calculations
\cite{Kolafa1992}, we choose to write this is as
\begin{align}
  u_j(\v x) &= \epsilon_{jlm} \left( \sum_{n=1}^N\frac{\v x- \v x^n}{|\v x- \v x^n|^3} f_l^n
  \right)_m .
\end{align}
Defining
\begin{align}
  \v F_l(\v x) = \sum_{n=1}^N\frac{\v x- \v x^n}{|\v x- \v x^n|^3} f_l^n,
  \label{eq:def_F}
\end{align}
we can write the potential as
\begin{align}
  u_j(\v x) &= \epsilon_{jlm} F_{lm}(\v x),
  \label{eq:u_from_F}
\end{align}
where $F_{lm}=(\v F_l)_m$. This means that we can use any method
available for electrostatic force computations to compute $\v
F_1$--$\v F_3$ at all target points, and then combine them as in
\eqref{eq:u_from_F} to get $\v u$.

\section{Ewald summation for the rotlet}

We now consider the case where we have $N$ source points contained in the box $L_1 \times
L_2 \times L_3$, which we will refer to as the primary cell. The periodic potential is
then defined as the potential from all source points in all periodic replications of the
primary cell,
\begin{align}
  u_j(\v x) &= \sum_{\v p\in\mathbb Z} \sum_{n=1}^N \Omega_{jl}(\v x + \v \tau(\v p) - \v
  x^n) f^n_l,
  \label{eq:periodic_sum}
\end{align}
where $\v\tau(\v p) = (L_1p_1, L_2p_2, L_3p_3)$ represents a periodic shift. The slow
decay of $\Omega$ makes this sum only conditionally convergent, which is why it is instead
computed using Ewald summation. For the electrostatic potential, the Ewald summation for
the periodic sum is \cite{Deserno1998}
\begin{align}
  \begin{split}
    \sum_{\v p\in\mathbb Z} \sum_{n=1}^N\frac{\v x + \v \tau(\v p) - \v x^n}{|\v x + \v
      \tau(\v p) - \v x^n|^3} q^n =
    \sum_{\v p\in\mathbb Z} \sum_{n=1}^N G^R(\v x + \v \tau(\v p) - \v x^n) q^n \\
    + \frac{4\pi i}{V}\sum_{\v k \ne 0} \frac{k_m}{k^2} e^{-k^2/4\xi^2} \sum_{n=1}^N q^n
    e^{-i\v k \cdot (\v x - \v x_n)},
  \end{split}
  \label{eq:ewald_force}
\end{align}
where
\begin{align}
  G^R(\v r,\xi) = \frac{\v r}{r^3}\left(\erfc(\xi r) + \frac{2 \xi r}{\sqrt{\pi}}
    e^{-\xi^2r^2} \right).
  \label{eq:G_force} % !
\end{align}
Here $V=L_1L_2L_3$ is the volume of the primary cell, and $k_i \in \{2\pi n/L_i : n \in
\mathbb Z\}$ are the Fourier space vectors. The first sum is called the real space sum; it
contains the short-range behavior of the kernel and converges rapidly in real space. The
second sum is called the Fourier space sum; it contains the long-range behavior of the
kernel and converges rapidly in Fourier space, due to its smoothness. The Ewald parameter
$\xi$ controls how short-range and smooth the two components are.

For the periodic rotlet potential \eqref{eq:periodic_sum}, we can make a similar
decomposition,
\begin{align}
  u_j(\v x) &= u_j^R(\v x) + u_j^F(\v x),
\end{align}
where $u^R$ is the real space sum and $u^F$ is the Fourier space sum,
\begin{align}
  u_j^R(\v x) &= \sum_{\v p\in\mathbb Z} \sum_{n=1}^N 
  \Omega^R_{jl}(\v x + \v \tau(\v p) - \v x^n, \xi) f^n_l ,
  \label{eq:real_sum} \\
  u_j^F(\v x) &= \frac{1}{V}\sum_{\v k \ne 0} 
  \widehat\Omega^F_{jl}(\v k, \xi) \sum_{n=1}^N f_l^n e^{-i\v k \cdot (\v x - \v x_n)} .
  \label{eq:fourier_sum}
\end{align}
Using \eqref{eq:def_F} and \eqref{eq:u_from_F}, we can identify the real and Fourier space
kernels from the Ewald decomposition of the electrostatic force
\eqref{eq:ewald_force}--\eqref{eq:G_force}, which gives us
\begin{align}
  \Omega^R_{jl}(\v r,\xi) &= \epsilon_{jlm}\frac{r_m}{r^3}\left(\erfc(\xi r) + \frac{2 \xi
      r}{\sqrt{\pi}} e^{-\xi^2r^2} \right),\\
  \widehat\Omega^F_{jl}(\v k,\xi) &= \epsilon_{jlm} 4\pi i\frac{k_m}{k^2} e^{-k^2/4\xi^2} .
\end{align}

\subsection{Zero wave number term}

The term corresponding to $\v k=0$ is omitted from the Fourier space sum
\eqref{eq:fourier_sum}, as $\widehat\Omega^F$ is singular at the origin. The term
corresponds to a constant ''ground level'' throughout the domain, and whether or not a
correction for this is required depends on the physics of the problem. For the
electrostatic potential no correction is required, which relates to the basic assumption
of charge neutrality \cite{Deserno1998}. In Stokes flow, a reasonable requirement is that
the periodic flow should have a zero mean. Denoting by $D_j$ the face of the primary cell
in the $x_j$-direction (lying in the plane $x_j=0$), the zero mean flow requirement can be
stated as
\begin{align}
  \avg{u_j} := \frac{1}{A_j} \int_{D_j} u_j(\v x) \d S(\v x) = 0 ,
  \label{eq:mean_flow_cond}
\end{align}
where $A_j=\int_{D_j}\d S(\v x)$.  For the stokeslet potential the $\v k=0$ term is zero,
and it is shown in \cite{Pozrikidis1996} that this is due to a balancing pressure gradient
in the direction of the point forces. For the stresslet potential the periodic sum does
generate a mean flow, and a correction term was derived in \cite{AfKlinteberg2014a} for
the case when the sum represents an integral over the surface of a rigid body.

To derive a result for the rotlet, we will now repeat the steps of the derivation in
\cite{AfKlinteberg2014a}. To that end, we will consider the periodic potential from a
point source of strength $\v f$ located at $\v x_s$. The Fourier transform of the periodic
sum \eqref{eq:periodic_sum} is then
\begin{align}
  u_j(\v x) = \frac{4\pi i}{V}\sum_{\v k \ne 0} \frac{k_m}{k^2} f_l e^{-i\v k \cdot (\v x
    - \v x_s)} + \widehat\Omega^0_{jl}f_l,
  \label{eq:pure_fourier_sum}
\end{align}
(this can be by seen by considering the limit $\xi\to\infty$ of the Ewald sum). Here
$\widehat\Omega^0$ is a correction for the $\v k=0$ term omitted in the sum. Inserting
\eqref{eq:pure_fourier_sum} into \eqref{eq:mean_flow_cond} and assuming no implicit
summation over $j$ in the following derivation, we get the requirement
\begin{align}
  \epsilon_{jlm} \frac{4\pi i}{V}\sum_{\v k \ne 0} \frac{k_m}{k^2} f_l 
  \int_{D_j} e^{-i\v k \cdot (\v x - \v x_s)} \d S(\v x)
  + A_j \widehat\Omega^0_{jl} f_l = 0 .
  \label{eq:k0_requirement}
\end{align}
The surface $D_j$ covers exactly one period in the directions perpendicular to $x_j$.
Hence, the integral is nonzero only if $k_i=0$ for $i \ne j$, such that
\begin{align}
  \sum_{\v k \ne 0} \frac{k_m}{k^2} \int_{D_j} e^{-i\v k \cdot (\v x - \v x_s)} \d S(\v x)
  = \delta_{mj} \sum_{k_j \ne 0} \frac{k_j}{k_j^2} A_j e^{-k_j (\v x_s)_j} .
\end{align}
Inserting this into \eqref{eq:k0_requirement}, we get that the correction term is zero,
\begin{align}
  \widehat\Omega^0_{jl} = - \epsilon_{jlj} \frac{4\pi i}{V} \sum_{k_j \ne 0}
  \frac{k_j}{k_j^2} A_j e^{-k_j (\v x_s)_j} = 0,
\end{align}
since $\epsilon_{ijk}=0$ if $i=k$. This means that the periodic rotlet sum produces zero
mean flow, and no correction term is needed in the Ewald summation.

\subsection{Self interaction}
When the target point $\v x$ in the periodic sum \eqref{eq:periodic_sum} is one of the
source points, i.e. $\v x=\v x^i$ for some $i\in[0,N]$, then the term corresponding to $\v
p=0$ and $n=i$ should be deleted from the summation, as it is singular. This is commonly
referred to as removing the self interaction of the point. 

When computing the periodic sum using Ewald summation, the part of the self interaction
that ends up in the real space sum is easy to remove, by simply omitting the corresponding
term in the summation. Part of the self interaction may however end up in the Fourier
space sum, in which case a correction term must be added (this is the case for the
electrostatic and stokeslet potentials \cite{Deserno1998,Lindbo2010}).

In the case of the rotlet, the self interaction correction turns out to be zero. One way
of seeing this is by considering the limit
\begin{align}
  \lim_{\v r\to 0} \left( \Omega(\v r)-\Omega^R(\v r) \right) = 0,
  \label{eq:r0_limit}
\end{align}
which can be shown by a series expansion of $\widehat\Omega^R$ around $\v r=0$. This means
that all of the self interaction is contained in the real space component, such that no
correction has to be added. Another way of seeing this is to consider the Fourier space
sum for the case of $N=1$,
\begin{align}
  u^F_j(\v x^1) = \frac{1}{V}\sum_{\v k \ne 0} \widehat\Omega^F_{jl}(\v k, \xi) f_l = 0,
\end{align}
since $\widehat\Omega^F$ is odd in $\v k$. This in turn implies \eqref{eq:r0_limit}.

\subsection{Final form}
Since no correction terms have to be added for self interaction or $\v k=0$, the final
form for the rotlet Ewald sum is as already stated,
\begin{align}
\begin{split}
  &\sum_{\v p\in\mathbb Z} \sum_{n=1}^N \Omega_{jl}(\v x + \v \tau(\v p) - \v x^n) f^n_l = \\
  &\sum_{\v p\in\mathbb Z} \sum_{n=1}^N \Omega^R_{jl}(\v x + \v \tau(\v p) - \v x^n, \xi)
  f^n_l
  + \frac{1}{V}\sum_{\v k \ne 0} \widehat\Omega^F_{jl}(\v k, \xi) \sum_{n=1}^N f_l^n
  e^{-i\v k \cdot (\v x - \v x_n)},
\end{split}
\label{eq:final_ewald_sum}
\end{align}
where
\begin{align}
  \Omega^R_{jl}(\v r,\xi) &= \epsilon_{jlm}\frac{r_m}{r^3}\left(\erfc(\xi r) + \frac{2 \xi
      r}{\sqrt{\pi}} e^{-\xi^2r^2} \right),
  \label{eq:rotlet_real}\\
  \widehat\Omega^F_{jl}(\v k,\xi) &= \epsilon_{jlm} 4\pi i\frac{k_m}{k^2} e^{-k^2/4\xi^2} .
  \label{eq:rotlet_k}
\end{align}

\section{Truncation errors}

\begin{figure}
  \centering
  \tikzset{mark size=1}
  \includegraphics[width=0.5\textwidth]{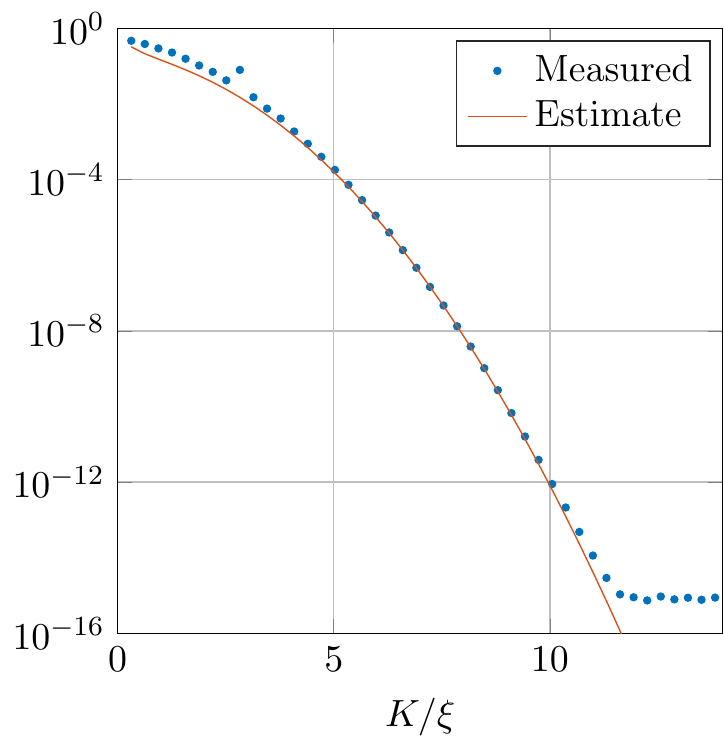}
  \caption{Fourier space RMS truncation error (relative) for $\xi=20$ and 1000 random
    sources in the unit cube. Estimate computed using \eqref{eq:est_fourier}.}
  \label{fig:k_error}
\end{figure}

\begin{figure}
  \centering
  \tikzset{mark size=1}
  \includegraphics[width=0.49\textwidth]{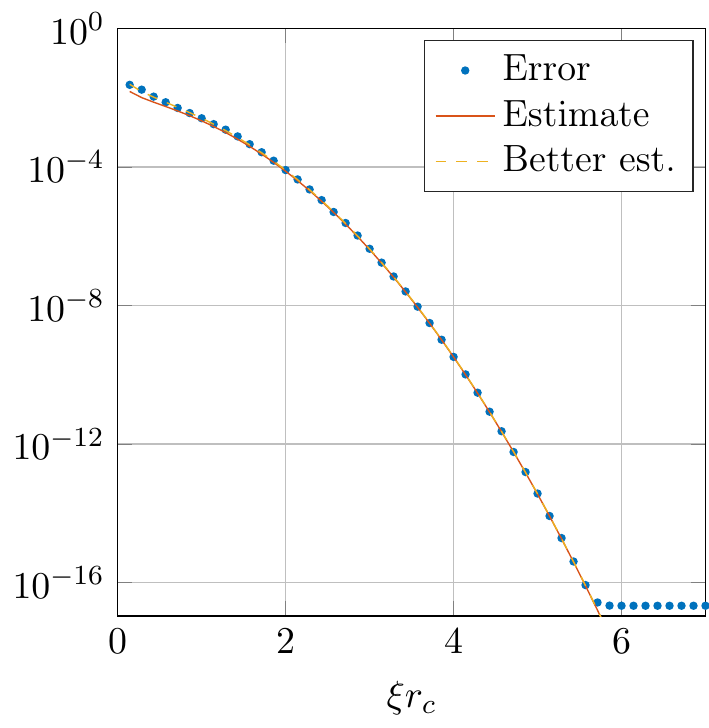}
  \includegraphics[width=0.49\textwidth]{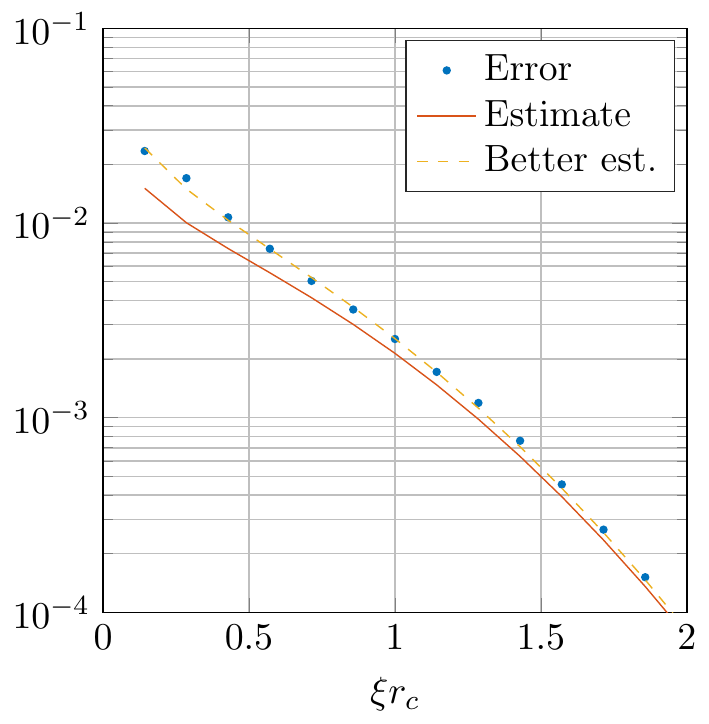}
  \caption{Real space RMS truncation error (relative) for $\xi=20$ and 1000 random sources
    in the unit cube. Estimate computed using \eqref{eq:est_real}, better estimate
    computed using \eqref{eq:rc_better}.}
  \label{fig:rc_error}
\end{figure}

When computing the Ewald sum \eqref{eq:final_ewald_sum} in practice, the real and Fourier
space sums must be truncated at some truncation radius $r_c$ and maximum wave number $K$,
such that
\begin{align}
  |\v x + \v \tau(\v p) - \v x^n| \le r_c \quad \text{and} \quad
  k \le K .
\end{align}
Estimates for the error committed when truncating the rotlet Ewald sum can be derived from
existing error estimates for the Ewald sum of the electrostatic force
\eqref{eq:ewald_force}. Let $\Delta \v F_l(\v x)$ be the error in a component $\v F_l(\v
x)$ \eqref{eq:def_F} when computing it using some numerical method (e.g. truncated Ewald
summation). The root mean square (RMS) error in $\v F_l$ can then be defined as
\begin{align}
  \delta \v F_l^2 = \frac{1}{N}\sum_{n=1}^N |\Delta \v F_l(\v x^n)|^2 .
  \label{eq:F_RMS}
\end{align}
This error can be approximated as
\begin{align}
  \delta \v F_l^2 \approx Q_l E,
\end{align}
where $E$ depends on the method and
\begin{align}
  Q_l = \sum_{n=1}^N (f_l^n)^2.
\end{align}
Based on \eqref{eq:u_from_F}, we now define
\begin{align}
  \delta \v u^2 = \frac{1}{N}\sum_{n=1}^N \sum_{j,l,m=1}^3 (\epsilon_{jlm} \Delta F_{lm}(\v
  x^n))^2 .
  \label{eq:def_du2}
\end{align}
Assuming the error to be equally distributed in all coordinate directions, we replace $\epsilon_{jlm}^2$ by its average
\begin{align}
  \overline{\epsilon^2_{jlm}} = \frac{1}{27}\sum_{j=1}^3 \sum_{l=1}^3 \sum_{m=1}^3
  \epsilon^2_{jlm} = \frac{2}{9},
  \label{eq:eps_avg}
\end{align}
such that, combining \eqref{eq:F_RMS}, \eqref{eq:def_du2} and \eqref{eq:eps_avg},
\begin{align}
  \delta \v u^2 \approx \frac{2}{9} \sum_{j,l=1}^3 \delta \v F_l^2 \approx
  \frac{2}{3} \sum_{l=1}^3 Q_l E = \frac{2}{3} Q E,
  \label{eq:u_RMS_from_E}
\end{align}
where
\begin{align}
  Q = \sum_{l=1}^3 Q_l = \sum_{n=1}^N |\v f^n|^2.
\end{align}
In the case of Ewald summation, a classic result by Kolafa \& Perram \cite{Kolafa1992}
gives a very accurate RMS error estimate for the electrostatic force potential, under the
assumption of randomly distributed sources and a Gaussian error distribution. The resulting
estimates for the real and Fourier space truncation errors are
\begin{align}
  E^R &= \frac{4}{V r_c} e^{-2\xi^2 r_c^2}, \\
  E^F &= \frac{4\xi^2}{\pi V K} e^{-K^2/2\xi^2} .
\end{align}
Together with \eqref{eq:u_RMS_from_E}, this gives us the error estimate for rotlet Ewald
sum:
\begin{align}
  \delta\v u = \delta\v u^R + \delta\v u^F,
  \label{eq:est}
\end{align}
where
\begin{align}
  \delta\v u^R &\approx \sqrt{\frac{8Q}{3V r_c}} e^{-\xi^2 r_c^2}, 
  \label{eq:est_real}\\
  \delta\v u^F &\approx \sqrt{\frac{8\xi^2Q}{3\pi V K}} e^{-K^2/4\xi^2} .
  \label{eq:est_fourier}
\end{align}
These estimates are very accurate, just like their electrostatic counterparts. Figures
\ref{fig:k_error} and \ref{fig:rc_error} show an example with $\xi=20$ and 1000 rotlet
point sources randomly distributed in the unit cube, with errors in real and Fourier space
computed by comparing to a converged reference solution. The estimates follow the measured
RMS errors extremely well, until full numerical precision is obtained around $K/\xi
\approx 12$ in Fourier space and $\xi r_c \approx 6$ in real space. These relations
actually give full numerical accuracy for a wide range of parameters, as the error
estimates are strongly dominated by their exponential terms.

The real space error estimate can be improved by explicitly evaluating the integral
estimated in \cite{Kolafa1992}. The resulting error estimate,
\begin{align}
  \delta\v u^R \approx \sqrt{\frac{8\pi Q}{3 V r_c} \left(
      \erfc(\xi r_c)^2 + \sqrt{\frac{2}{\pi}} \xi r_c \erfc(\sqrt{2} \xi r_c)
    \right)},
  \label{eq:rc_better}
\end{align}
follows the measured RMS error estimate more closely also for small $\xi r_c$ (''Better
estimate'' in Figure \ref{fig:rc_error}).  In practice the difference might however not be
significant enough to merit using the more cumbersome expression.

\section{Concluding remarks}

By making use of the correspondence between the rotlet and the kernel for the
electrostatic force, we have derived an Ewald summation for the periodic rotlet potential
\eqref{eq:final_ewald_sum}--\eqref{eq:rotlet_k}, as well as accurate truncation error
estimates \eqref{eq:est}--\eqref{eq:est_fourier} for the Ewald sum. Coupled with a fast
Ewald summation method, such as the spectral Ewald method \cite{Lindbo2010}, these results
allow the periodic rotlet potential to be computed rapidly and with controlled precision.

\section{Supplementary material}

The Ewald decomposition for the rotlet described in this text has been implemented in the
Spectral Ewald package, which is available as open source software at
\url{http://github.com/ludvigak/SE_unified} . The package includes a script
(\texttt{SE\_Rotlet/demo.m}) that generates the plots of Figures \ref{fig:k_error} and
\ref{fig:rc_error}.

\bibliographystyle{jabbrv_abbrv_doi}
\bibliography{library}

\end{document}